\documentclass[12pt]{article}
\def\beqan{\begin{eqnarray*}}
\def\eeqan{\end{eqnarray*}}
\def\beqa{\begin{eqnarray}}
\def\eeqa{\end{eqnarray}}
\def\beq{\begin{equation}}
\def\eeq{\end{equation}}
\def\fracp#1#2{\frac{\partial#1}{\partial#2}}
\def\non{\nonumber}
\def\L{\left}
\def\R{\right}

\def\ty{\tilde{y}}
\def\tx{\tilde{x}}
\def\AdStS{$AdS_5\times S^5$ }
\def\I{{\rm I}}
\def\II{{\rm I\!I}}
\def\III{{\rm I\!I\!I}}
\def\approxx{\equiv}
\def\rat{y}
\def\IIsII{\II\ \!\!\II}

\def\lambdas{\lambda}
\def\mus{\mu}
\def\nus{\nu}

\makeatletter

\begin{document}
\renewcommand{\thefootnote}{\fnsymbol{footnote}}
\begin{titlepage}
\begin{flushright}
UT-KOMABA/04-12\\
October 2004
\end{flushright}
\begin{center}

{\bf\Large
Holography at string field theory level:

Conformal three point functions of BMN operators
}
\vspace{1.5cm}

{\bf  Hidehiko~Shimada}\footnote
{
E-mail: shimada@hep1.c.u-tokyo.ac.jp
}	

\vspace{1.0cm}

{\it  Institute of Physics, Tokyo University,\\
Komaba, Megro-ku, Tokyo 153-8902, Japan}

\vspace{1.9cm}
\end{center}
\begin{abstract}
A general framework for applying the pp-wave approximation
to holographic calculations in the AdS/CFT correspondence is proposed.
By assuming the existence and some properties of string field
theory (SFT) on $AdS_5 \times S^5$ background,
we extend the holographic ansatz proposed by
Gubser, Klebanov, Polyakov and Witten to SFT level. 
We extract relevant information of assumed SFT on $AdS_5 \times S^5$
from its approximation, pp-wave SFT.
As an explicit example, we perform 
string theoretic calculations of
the conformal three point functions of the BMN operators.
The results agree with the previous calculations in gauge theory. 
We identify a broad class of field redefinitions,
including known ambiguities of the interaction Hamiltonian,
which does not affect the 
results.
\end{abstract}
\vfill
\end{titlepage}
\setcounter{footnote}{0}
\renewcommand{\thefootnote}{\arabic{footnote}}

The AdS/CFT correspondence conjecture \cite{RBAdSCFT}, which states
that string theory on \AdStS is
equivalent to $\mathcal{N}=4$ 
supersymmetric Yang-Mills theory, 
is one of the most explicit proposal of 
equivalence between 
large $N$ gauge theory and string theory.
A characteristic feature of the correspondence is
the holography\cite{RBGKP}\cite{RBWitten}:
observables of gauge theory
seem to be related to
the behaviour of
string theory at the boundary of $AdS$ space.

An important problem is to understand the fundamental mechanism of
the correspondence, 
in other words, to understand how degrees of freedom of closed string 
arise from those of gauge theory.
Solving this problem will considerably improve our understanding of
string theory.
As a first step to 
understand the mechanism, 
it will be useful to 
calculate corresponding observables in
both string theory and gauge theory independently 
and check the agreement.

However, as is well known, there are difficulties in general to carry out
such comparisons.
At the string theory side, quantum string theory 
on \AdStS is not defined so that we cannot treat
non-zero modes of closed strings.
At the gauge theory side,
we have no general methods of computation since
perturbative methods do not work, the expansion parameter
$g_{\rm YM}^2 N$ being large.

A breakthrough has been made in 
\cite{RBBMN}.
At the string theory side, the pp-wave approximation is found to be very useful
for states 
which have large orbital angular
momentum $J$ on $S^5$.
Quantum theory is well defined under the approximation,
so that in particular one can treat non-zero modes of closed strings.
At the gauge theory side,
corresponding operators (BMN operators) are proposed.
The expansion parameter becomes $\frac{g_{\rm YM}^2 N}{J^2}$ for them.
Hence, perturbative methods are effective
in the regime $\frac{g_{\rm YM}^2 N}{J^2} <<1$.

By these methods, many tests are performed
based on the postulated equivalence between the energy in string theory
and the dilatation operator in gauge theory.
However, holographic aspects of the AdS/CFT correspondence 
via the pp-wave approximation
remain much unexplored.
A step was made towards this direction by the work by Dobashi, the author,
Yoneya\cite{RBDSY}.
The main observation is that,
in order to directly apply the pp-wave approximation
to the holography, we should interpret 
the closed strings as
in a state of tunnelling under a barrier
of the gravitational potential in $AdS$ background.

In this letter
we propose, under the tunnelling picture, a general framework to
apply the pp-wave approximation to holographic 
evaluation of conformal $n$-point correlation functions of BMN operators.
We in particular perform explicit calculations
for the three point functions
of scalar BMN operators to the leading order. 
We find agreement with the 
previous calculation in gauge theory
up to an overall factor.
Our methods do not suffer from
known ambiguities in the pp-wave SFT Hamiltonian.
We identify a broad class of field redefinitions
including them
which does not affect the physical observable.

The basic strategy is as follows.
We begin by assuming the existence of
string field theory (SFT) on the \AdStS background.
Then we show that there is a very straightforward
extension,
from supergravity level to SFT level,
of the holographic ansatz given in \cite{RBGKP}\cite{RBWitten}.
Full construction of SFT on \AdStS would be a hard task.
However, we can extract relevant information 
of SFT on \AdStS
from pp-wave SFT
which can be considered as an approximation of the near BPS sector
of it.
We introduce coordinates  and 
basis (see (\ref{RFpplikecooridnate}),(\ref{RFnegbas}), (\ref{RFposbas}) below)
which facilitate
the extraction of information.
We obtain a novel representation (\ref{RFTheRule}) 
of the observable in gauge theory 
by an infinite series of matrix elements of SFT on \AdStS,
which reduces to (\ref{RFTheRulepp})
under the pp-wave approximation.

Let us start by making minimum assumptions
on the nature of SFT on \AdStS.
Firstly, we assume that the 
string field consists of infinitely many 
fields on $AdS_5$.
Secondly, we assume that 
the free part of the 
SFT action is given by the usual 
Klein-Gordon operators on $AdS_5$, 
at least for scalar fields (on $AdS_5$) in the string field.

The two assumptions enable us to propose the following
extension of the holographic ansatz.
Firstly, we consider a correspondence between
fields $\phi_L$ in SFT on \AdStS  and 
non-BPS composite operators $\mathcal{O}_L$
(with definite conformal dimensions)
in gauge theory,
extending the correspondence between 
fields in supergravity theory and BPS operators.
The extend ansatz is then given by 
\footnote{
Actually, the left hand side of the equation should be
considered as
a tree level approximation (in string theory)
of the path integral of $e^{-S}$.
}
\beq
<e^{-\int J_L(x) \cdot \mathcal{O}_L(x) d^4\!\!x} >
= e^{-S[\phi_{cl}]}, \label{RFholo}
\eeq
where $\phi_{cl}$ is the classical solution of
the equation of motion of SFT such that the 
asymptotic behaviour near the  boundary (i.e. $z\approx0$) is
given by $\phi_{L}\approx z^{4-\Delta_L} J_L(x)$
and $S$ is the action of SFT.
The asymptotic behaviour follows from the
Klein-Gordon equation, so it is the same as the supergravity case.
We work on the Euclideanized $AdS_5$
and use the Poincar\' e coordinates,
$ds^2=\frac{R^2 (dz^2+ (dx^\mu)^2)}{z^2}$.
$R$ is the radius of $AdS_5$, which is related to gauge theory
by $R^4/\alpha^{\prime 2}\sim g_{\rm YM}^2 N$. 
We choose the length scale so that $R$ is equal to unity.
Indices $\mu, \nu,\ldots$ run from $0$ to $3$.
$\Delta_L$ is the conformal dimension of $\mathcal{O}_L$
which is related to the mass $m_L$ of $\phi_L$
by $\Delta_L=2+\sqrt{4+m_L^2}$.
The important advantage of the original ansatz is 
that the $n$-point functions given by it 
automatically satisfy the correct conformal transformation law.
This property is preserved upon our extension.

For the near-BPS sector, we can determine
$\Delta$, appearing in the asymptotic behaviour
near boundary, by the BMN formula
\beq
\Delta-J=\sum N_m \sqrt{1+\frac{R^4 m^2}{J^2 \alpha^{\prime 2}}},
\eeq
which holds approximately when J is large.
Thus we have been able to extract the masses of the fields,
the only information
of the free part of the SFT action, in a very simple way.

The interaction part, to which we now turn, requires much more
work. One reason is that the interaction has strongly non-local
features: we cannot assume the local interaction form 
as had been done, for example in \cite{RBhololocalSG},
in the supergravity case.
From now on, we shall concentrate on three point functions.
Let us first identify what we should calculate
according to the holographic ansatz.
We have three scalar BMN operators,
$\mathcal{O}_r (r=1, 2, 3)$, and 
corresponding scalar fields on $AdS_5$, $\phi_r$.
The holographic ansatz reads, for three point functions,
\beq
\langle\mathcal{O}_1(x_1)\mathcal{O}_2(x_2)\mathcal{O}_3(x_3)\rangle
=S_{int}[K_1,K_2,K_3], \label{RFholo3}
\eeq
where $S_{int}$ is the cubic interaction part of the SFT action
and $K_r$'s are the boundary-bulk propagators  for the fields $\phi_r$ 
given by
\footnote{
The prefactor 
depends only on $\Delta_r$. 
We neglect it in this letter.
By taking care of overall factors,
comparison to works 
\cite{RBGomisMoriyamaPark} should be possible.
We defer the study to further publication.
}
\beq
K_r= 
\frac{\Gamma(\Delta_r)}{\sqrt{\pi}^4 \Gamma(\Delta_r-2)} \L(\frac{z}{z^2+(x-x_r)^2}\R)^{\Delta_r}.
\label{RFbbprop}
\eeq
Thus, what we want is the value of the interaction action
evaluated on some special field 
configurations,  namely boundary-bulk propagators.

On the other hands, 
what we have is the three string vertex,
which embodies the joining (or splitting) amplitudes
of strings.
It gives roughly the value of 
coefficients before  the $A^\dagger A A$ (or $A^\dagger A^\dagger A$)
terms in the interaction Hamiltonian,
where $A, A^\dagger$ respectively are the
annihilation, creation operators of the closed string
(in the second-quantized sense).
The interaction action is given by the time integral of 
the interaction Lagrangian,
which in turn is the 
Legendre transformation of the interaction Hamiltonian.
Hence, we should be able to calculate the interaction 
action from the three string vertex.
However, we should clarify some points.
Firstly, we have to understand what is meant by the time in our case.
Moreover, we should decide 
the annihilation-creation operators,
or the basis of the positive and negative energy solutions
to the equations of motion of the free fields,
to use.
We shall specify the basis 
and then expand the boundary-bulk propagators by them 
below.

In flat space, it is most natural to take the basis of the form 
like $e^{i k x}$. However, we are working in $AdS$ space so that
there is an external potential.
Since it has the harmonic oscillator form under pp-wave approximation,
the natural choice made in pp-wave SFT is 
the wavefunctions for the harmonic oscillator:
the wavefunction given by a gaussian (the ground state) and gaussian multiplied by
hermite polynomials (excited states).
We shall seek for the exact basis which reduce to these
gaussian wavefunctions under the pp-wave approximation.
We will see that
the discussion crucially depends on
the use of the exact basis.

To this end, we introduce new coordinates $\ty, \tx^\mu$ on $AdS_5$,
which capture qualitative features of the pp-wave background,
\beq
\tilde{y}= \log{\sqrt{z^2+(x^\mu)^2}},\quad \tilde{x^\mu}=\frac{x^\mu}{z}.
\label{RFpplikecooridnate}
\eeq
The metric becomes
\beq
ds^2= (1+ \tilde{x}^2) d\tilde{y}^2 + (d\tilde{x}^\mu)^2 
- \frac{(\tilde{x}^\mu d \tilde{x}^\mu)^2}{1+\tx^2}.
\eeq
We emphasize that this is only a coordinate transformation
so that there are  no approximations involved. 
We are still quantitatively  working in $AdS_5$.
The isometry 
$z^\prime =\alpha z, x^\prime =\alpha x$,
which corresponds to the dilatation transformation
in gauge theory, is realized by the translation of $\tilde{y}$,
$\ty^\prime=\ty+\log\alpha$.
Also, the isometry of $AdS_5$ corresponding to inversion
\begin{equation}
z^\prime=\frac{z}{z^2+x^2},\quad
x^{\prime\mu}=\frac{x^\mu}{z^2+x^2}, \label{RFinversion}
\end{equation}
is realized as the $\tilde{y}$-reversal, $\ty^\prime=-\ty$.

We consider $\tilde{y}$ as Euclidean time. 
This identification follows naturally from
the tunnelling picture of \cite{RBDSY},
and enables us to directly identify the energy in string theory and
the dilatation operator
in gauge theory.

Let us consider the well-known solution 
to the Klein-Gordon equation $z^\Delta$, in the new coordinates,
\beq
z^\Delta= \L(\frac{1}{\sqrt{1+\tx^2}}\R)^\Delta e^{\Delta\ty}. \label{RFgroundstatewf}
\eeq
For large $\Delta$, we can approximate the lorentzian 
in the above expression by a gaussian,
\beq
\L(\frac{1}{\sqrt{1+\tx^2}}\R)^\Delta \approx e^{-\frac{\Delta}{2} \tx^2}.
\eeq
Thus, we have found the exact solution which reduces to the 
gaussian wavefunction under the pp-wave approximation, large
J implying large $\Delta$.
\footnote{
Another well-known solution 
$z^{4-\Delta}$ corresponds to a non-normalisable 
wavefunction, $e^{+\frac{\Delta}{2} \tx^2} e^{-\Delta \ty}$.
}
From the dependence on $\ty$, we see that 
it is the negative energy solution (in the sense of Euclidean field theory)
of energy $\Delta$.
By inversion, we get the corresponding positive energy solution
\beq
(z^\prime)^\Delta=\L(\frac{z}{z^2+x^2}\R)^\Delta
=\L(\frac{1}{\sqrt{1+\tx^2}}\R)^\Delta e^{-\Delta \ty}.
\eeq

We next seek for the solutions 
which reduce to the wavefunctions 
given by gaussian multiplied by hermite polynomials.
To this end, let us consider the Killing vector field
\begin{equation}
\L(\fracp{}{x^{\prime\mu}}\R)_{z^\prime}=
e^{+\ty} \L(\sqrt{1+\tx^2}\fracp{}{\tx^\mu}
-\frac{\tx^\mu}{\sqrt{1+\tx^2}} \fracp{}{\ty}\R), \label{RFladderAdS}
\end{equation}
which corresponds to the special conformal symmetry in gauge theory.
Being Killing vector field, it gives a new solution to the Klein-Gordon
equation when it acts upon a solution.
Furthermore, since 
wavefunctions have essentially the gaussian form $e^{-\frac{\Delta}{2} \tx^2}$,
$\tx$ should be considered 
as  a small quantity
of order $\frac{1}{\sqrt{\Delta}}$. 
Also, we have $\fracp{}{\ty} \approx \mp \Delta$, 
the sign reflecting whether we apply the
operator to positive or negative frequency solutions. 
By using these approximations
we get
\beq
\L(\fracp{}{x^{\prime\mu}}\R)_{z^\prime}\approx
e^{+\ty} \sqrt{2 \Delta}\L( \sqrt{\frac{1}{2 \Delta}}\fracp{}{\tx^\mu} \label{RFladderAdSapprox}
\pm \sqrt{\frac{\Delta}{2}}\tx^\mu\R).
\eeq
The right hand side is, apart from the factor $\sqrt{2 \Delta}$, 
the ladder operator (or the annihilation, creation
operator in the first quantized sense) of the closed string zero modes
in pp-wave.~\footnote{
The factor $e^{+\ty}$ arises because of 
representing the operator in the Heisenberg picture.
}
Therefore, we can construct the desired wavefunctions
by applying (\ref{RFladderAdS}) several times
to the ground state wavefunction (\ref{RFgroundstatewf}).
There is also the inverted version $\L(\fracp{}{x^{\mu}}\R)_{z}$, 
corresponding to the translational symmetry in gauge theory.

Thus we have obtained the basis of the 
negative and positive energy solutions
\beqa
\L(\fracp{}{x^{\prime\mu_1}}\R)_{z^\prime}
\ldots
\L(\fracp{}{x^{\prime\mu_n}}\R)_{z^\prime}
z^\Delta
&=&
\Psi_{\mu_1 \ldots \mu_n}(\tx) e^{(\Delta+n) \ty}, \label{RFnegbas}
\\
\L(\fracp{}{x^{\mu_1}}\R)_{z}
\ldots
\L(\fracp{}{x^{\mu_n}}\R)_{z}
(z^\prime)^\Delta
&=&
\Psi_{\mu_1 \ldots \mu_n}(\tx) e^{-(\Delta+n) \ty}, \label{RFposbas}
\eeqa
respectively, where 
\beq
\Psi_{\mu_1 \ldots \mu_n}(\tx) \approx
\sqrt{2 \Delta}^n a_0^{\mu_1 \dagger} \ldots a_0^{\mu_n \dagger} 
e^{-\frac{\Delta}{2} \tx^2}. \label{RFbasis2ladderFactor}
\eeq
Here,
$a_0^{\mu \dagger}$'s
denote the creation operators 
of the closed string zero modes which have polarisations, labeled by $\mu$'s,
corresponding to insertion of vector impurities to BMN operators.

Our next task is to 
expand the boundary-bulk propagator (\ref{RFbbprop}) by the basis.
It is readily seen that the expansion of (\ref{RFbbprop}) by $x_r$
gives just the expansion by the basis,
\beq
\L(\frac{z}{z^2+(x-x_r)^2}\R)^{\Delta_r}=
\sum_n \frac{1}{n!} (-x_r^{\mu_1})\ldots(-x_r^{\mu_n})
\L(\fracp{}{x^{\mu_n}}\R)_z \ldots \L(\fracp{}{x^{\mu_1}}\R)_z
\L(\frac{z}{z^2+x^2}\R)^{\Delta_r}.
\label{RFexpfuture}
\eeq
In this expansion only positive energy solutions,
which are exponentially decreasing, appear.
Hence, it is well-behaved when $\ty \rightarrow +\infty$.
On the other hand, 
it seems that the propagator becomes singular
when $\ty \rightarrow -\infty$, at first sight.
Actually, this is not the case.
Interestingly, the expansion above converges
only in certain region, which is given by $\ty>\log{|x_r|}$. 
In the opposite region $\ty<\log{|x_r|}$, the propagator has another expansion.
We first write the propagator
in the inverted frame, and then perform similar expansion. The result is 
\beq
\L(\frac{z}{z^2+(x-x_r)^2}\R)^{\Delta_r}=
\frac{1}{|x_r|^{2 \Delta_r}}\sum_n \frac{1}{n!} 
\L(-\frac{x_r^{\mu_1}}{|x_r|^2}\R)\ldots\L(-\frac{x_r^{\mu_n}}{|x_r|^2}\R)
\L(\fracp{}{x^{\prime\mu_n}}\R)_{z^\prime} \ldots \L(\fracp{}{x^{\prime\mu_1}}\R)_{z^\prime} 
z^{\Delta_r},
\label{RFexppast}
\eeq
in which only exponentially increasing solutions appear.
Thus the expansion of the propagator is actually quite well-behaved.
The existence of the critical time, $\log|x_r|$, is essential to the
following calculations. The origin of the
circle of convergence is that the propagator is a
rational function.
We would not see the existence of the critical time
if we used the approximate gaussian wavefunctions
instead of the exact basis.

We now compute the right hand side of the holographic 
ansatz (\ref{RFholo3}) by the expansions (\ref{RFexpfuture}) and
(\ref{RFexppast}).
There are critical time 
$\log{|x_1|}$, $\log{|x_2|}$, $\log{|x_3|}$,
corresponding to
the three scalar fields. 
We hereafter fix the radial order to be
$|x_1|<|x_2|<|x_3|$.
Then we have four regions,
$-\infty<\ty<\log{|x_1|}$,
$\log{|x_1|}<\ty<\log{|x_2|}$,
$\log{|x_2|}<\ty<\log{|x_3|}$,
$\log{|x_3|}<\ty<\infty$.
In each of these regions, we have expansion
labeled by three integers corresponding to the three fields.
In each term in the expansion, the 
scalar fields behave exponentially in $\ty$.
Since $S_{int}$ is linear in each of the fields,
the integrand itself is an exponential function.
We define the matrix elements
$L$'s
between the members of the basis (\ref{RFnegbas})(\ref{RFposbas}) by
\beqa
&&S_{int}\L[
\Psi_{\lambda_1 \ldots \lambda_l} e^{\pm_1 (\Delta_1+l) \ty},
\Psi_{\mu_1 \ldots \mu_m} e^{\pm_2 (\Delta_2+m) \ty},
\Psi_{\nu_1 \ldots \nu_n} e^{\pm_3 (\Delta_3+n) \ty}
\R]
\non\\
&=&L^{\pm_1\pm_2\pm_3}_{\lambda_1 \ldots \lambda_l:
\mu_1 \ldots \mu_m:
\nu_1 \ldots \nu_n} 
e^{(\pm_1(\Delta_1+l) 
\pm_2(\Delta_2+m) 
\pm_3(\Delta_3+n))\ty}. \label{RFdefmatrixelements}
\eeqa
The signs $(\pm)_r$ for the fields $\phi_r$ reflect whether
the negative or positive energy solutions are considered.
Then, by performing the integral in each region, we obtain,
\beqa
&&<\mathcal{O}_1(x_1)\mathcal{O}_2(x_2)\mathcal{O}_3(x_3)>\non\\
&&\approxx\sum_{l,m,n=0}^{\infty} \frac{(-1)^{l+m+n}}{l!m!n!}
\L(\frac{x_1^{\lambda_1}}{|x_1|}\ldots\frac{x_1^{\lambda_l}}{|x_1|}\R)
\L(\frac{x_2^{\mu_1}}{|x_2|}\ldots\frac{x_2^{\mu_m}}{|x_2|}\R)
\L(\frac{x_3^{\nu_1}}{|x_3|}\ldots\frac{x_3^{\nu_n}}{|x_3|}\R)\non\\
&&
\Bigg[\L(
-M^{---}_{\lambdas:\mus:\nus}
+M^{--+}_{\lambdas:\mus:\nus}
\R)
|x_3|^{-\Delta_1-\Delta_2-\Delta_3}
\L(\frac{|x_1|}{|x_2|}\R)^{l}\L(\frac{|x_2|}{|x_3|}\R)^{l+m}
\non\\ 
&&
+\L(
-M^{--+}_{\lambdas:\mus:\nus}
+M^{-++}_{\lambdas:\mus:\nus}
\R)
|x_2|^{-\Delta_1-\Delta_2+\Delta_3}|x_3|^{-2\Delta_3}
\L(\frac{|x_1|}{|x_2|}\R)^{l}\L(\frac{|x_2|}{|x_3|}\R)^{n}
\non\\
&&
+\L(
-M^{-++}_{\lambdas:\mus:\nus}
+M^{+++}_{\lambdas:\mus:\nus}
\R)
|x_1|^{-\Delta_1+\Delta_2+\Delta_3}|x_2|^{-2\Delta_2}|x_3|^{-2\Delta_3}
\L(\frac{|x_1|}{|x_2|}\R)^{m+n}\L(\frac{|x_2|}{|x_3|}\R)^{n}
\Bigg], \label{RFfullexpansionS}
\eeqa
where, we have introduced an abbreviation
\[
\frac{L^{\pm_1 \pm_2 \pm_3 }_{
\lambda_1 \ldots \lambda_l:
\mu_1 \ldots \mu_m:
\nu_1 \ldots \nu_n}}{
\pm_1 (\Delta_1+ l)\pm_2 (\Delta_2+ m)\pm_3 (\Delta_3+ n)}
=M^{\pm_1 \pm_2 \pm_3 }_{\lambdas:\mus:\nus}.
\]
We use $\approxx$
to signify that the equality holds up to a $\Delta$-dependent overall factor.
Three terms come from critical time 
$\log{|x_3|}$, $\log{|x_2|}$, $\log{|x_1|}$, respectively.
They do not mix in general
since the differences of $\Delta$'s are non-integral for generic operators.

On the other hands, dependence of the three point function on $x_r$ is 
fixed by conformal symmetry to be,
\beq
\frac{C}{
|x_1-x_2|^{\Delta_1 + \Delta_2 -\Delta_3}
|x_2-x_3|^{\Delta_2 + \Delta_3 -\Delta_1}
|x_3-x_1|^{\Delta_3 + \Delta_1 -\Delta_2}
}, \label{RF3pf}
\eeq
where $C$ is a constant.
The expansion (\ref{RFfullexpansionS})
should agree with the expansion of 
(\ref{RF3pf}) by the two parameters 
$\frac{|x_1|}{|x_2|}<1$ and $\frac{|x_2|}{|x_3|}<1$,
provided that conformal symmetry 
is properly realized in the interaction action i. e.
in the matrix elements $L$'s.
We can read off many identities embodying the conformal symmetry
of the matrix elements by fully comparing the two expansions.
In particular,
only the second term should be non-vanishing in (\ref{RFfullexpansionS}).
By exploiting these identities one may gain  
some insights on SFT on \AdStS.
In this letter, we shall instead concentrate on
deriving the factor $C$, by
comparing
the leading term in the expansion of (\ref{RF3pf})
and the $l=0$, $n=0$ part of the
second term in (\ref{RFfullexpansionS}).
We obtain,
\beq
C\approxx\sum_{m=0}^\infty \frac{(-1)^m}{m!}
\frac{x_2^{\mu_1}}{|x_2|}\ldots \frac{x_2^{\mu_m}}{|x_2|}
\L(
-\frac{L_{:\mu_1\ldots\mu_m:}^{--+}
}{-\Delta_1-(\Delta_2+m)+\Delta_3}
+ 
\frac{L_{:\mu_1\ldots\mu_m:}^{-++}
}{-\Delta_1+(\Delta_2+m)+\Delta_3}
\R).
\label{RFTheRule}
\eeq
This formula is our main result.
It is  interesting that 
the simple observable $C$ should be written by an infinite series
of the matrix elements with excited zero-modes.
We emphasize that 
no approximations (such as the pp-wave approximation)
are involved in this expression.

A couple of comments are in order.
Firstly, we have performed a consistency check for this expression
using toy models
which have local interactions.
The matrix elements $L_{:\mu_1\ldots\mu_m:}^{-\pm+}$ are calculated 
and then the series is evaluated.
The results agree with those which are obtained by direct integration
over $z, x$,
although we do not give explicit formulae here.
Secondly, 
we see that there may be some subtlety for BPS operators since 
the denominators may vanish for some $m$ due to their integral
$\Delta$'s.

From now on, we shall consider the application of (\ref{RFTheRule}).
To be specific, we present the calculations 
for the BMN operators 
\footnote{
Here, we refer to 
those mixed with double trace operators
to have definite conformal dimensions.
}
$\mathcal{O}_{\I},
\mathcal{O}_{\II}, \mathcal{O}_{\III}$ corresponding to the 
states in string theory,
\beqa
|\mathcal{O}_{\I} \rangle=a_{\I\ m_\I}^{\alpha_\I \dagger}
                    a_{\I -m_\I}^{\beta_\I \dagger} |0;J_\I\rangle, \quad
|\mathcal{O}_{\II} \rangle=|0;J_{\II}\rangle,\quad
|\mathcal{O}_{\III} \rangle=a_{\III m_{\III}}^{\alpha_\III \dagger}
                      a_{\III -m_{\III}}^{\beta_\III \dagger}|0;J_{\III}\rangle,
\eeqa
respectively. 
These states satisfy the level matching condition.
$|0;J_{r}\rangle (r=\I, \II, \III)$ denote the first-quantised
vacuum states of the closed strings with angular momentum $J_{r}>0$ satisfying
$J_{\I}+J_{\II}=J_{\III}$.
$a^\dagger_{r m}$'s denote creation operators 
of the $m$-th mode of the $r$-th string.
$\alpha$'s and $\beta$'s take one of the 
four values  corresponding to insertion of
scalar impurities.
We take $m_\I, m_\III >0$.
We put operators $\mathcal{O}_{\I},
\mathcal{O}_{\II}, \bar{\mathcal{O}}_{\III}$ respectively at the points 
$x_1, x_2, x_3$ (satisfying $|x_1|<|x_2|<|x_3|$).
\footnote{
There are six possibilities regarding the radial order of
the points where $\mathcal{O}_{\I},
\mathcal{O}_{\II}, \bar{\mathcal{O}}_{\III}$ are inserted.
All discussions work the same if we put 
the operators $(\mathcal{O}_{\I}, \mathcal{O}_{\II}, \bar{\mathcal{O}}_{\III})$ on 
$(x_3, x_2, x_1), (x_2, x_1, x_3), (x_3, x_1, x_2)$, respectively.
However, if we put them on $(x_1, x_3, x_2), (x_2, x_3, x_1)$,
both terms in (\ref{RFTheRule}) involve particles with negative $J$.
In order to treat these terms, we should carefully work out
the transformation law between matrix elements
in lightcone frame and in ordinary temporal frame.
This issue will be discussed elsewhere.
}

By angular momentum conservation, 
the second term in (\ref{RFTheRule}) involves particles which have negative $J$. 
The contribution of them should be negligible,
as is usual in the physics of the infinite momentum frame.
For the first term, 
we substitute the matrix elements $L$ defined via the Lagrangian
by the matrix elements of the Hamiltonian,
since Legendre transformation
between them 
involves only an
overall factor, at least in the leading order.
Then (\ref{RFTheRule}) becomes, using (\ref{RFbasis2ladderFactor}), 
\beq
C\approxx\sum_{n=0}^{\infty} \frac{(-1)^n}{n!} 
\frac{x_2^{\mu_1}}{|x_2|} \ldots \frac{x_2^{\mu_n}}{|x_2|}
\frac{
\langle\mathcal{O}_{\I}, \mathcal{O}_{\II}, \bar{\mathcal{O}}_{\III}|
\sqrt{2\Delta}^n a_{\II 0}^{\mu_1} \ldots a_{\II 0}^{\mu_n} 
|V\rangle}{\Delta_\I+(\Delta_{\II}+n)-\Delta_{\III}},
\label{RFTheRulepp}
\eeq
where $|V\rangle$ denotes the three string vertex.

For the class of 
interaction vertices with prefactors
quadratic in $a^\dagger$, such as those given in \cite{RBSpVol} or 
\cite{RBRedefVertex},
the matrix elements can be manipulated 
as follows defining quantities $E$ and $F$,
\beq
\langle \mathcal{O}_{\I}, \mathcal{O}_{\II}, \bar{\mathcal{O}}_{\III}|
a^{\mu_1}_{\II 0}\ldots a^{\mu_n}_{\II 0}
|V\rangle
=
\langle0|
a^{\mu_1}_{\II 0}\ldots a^{\mu_n}_{\II 0}
(E + F N^{\IIsII}_{00}
a^{\mu \dagger}_{\II 0} a^{\mu \dagger}_{\II 0}) 
e^{\frac{1}{2}a_{\II 0}^{\nu \dagger} N^{\IIsII}_{0 0} a_{\II 0}^{\nu \dagger}}
|0 \rangle,
\eeq
where $N^{\IIsII}_{0 0}$ is a Neumann coefficient. 
$E$ is the matrix element without any zero mode insertions,
while $F$ comes from zero modes in the prefactor.
Substituting into (\ref{RFTheRulepp}), we get
\beq
C\approxx
\sum_{\frac{n}{2}=0, 1, \cdots}
\frac{1}{\frac{n}{2}!}
\frac{E}{\Delta_\I+(\Delta_\II+n)-\Delta_\III}
(\Delta N^{\IIsII}_{00})^{\frac{n}{2}}
+
\sum_{\frac{n}{2}=1, 2, \cdots}
\frac{1}{\frac{n}{2}!}
\frac{n F}{\Delta_\I+(\Delta_\II+n)-\Delta_\III}
(\Delta N^{\IIsII}_{00})^\frac{n}{2}. \label{RFCisEFseries}
\eeq

Let us show the validity of using the pp-wave approximation
in the above series.
The validity holds
if the fluctuation $\langle x\rangle$ of the
oscillator is sufficiently small compared to the radius of the AdS
space, $\langle x\rangle <\!\!< R$. In terms of the excitation number 
$n$ in (\ref{RFCisEFseries}), this condition reads $n<\!\!<J$. 
Therefore, we should check that the leading contributions to the
series should come from the terms satisfying $n<\!\!<J$.
Since we wish to compare the results to perturbative calculation in gauge theory,
we shall work in the regime $\frac{g_{\rm YM}^2 N}{J^2} <\!\!<1$.
Since 
$N^{\IIsII}_{00}$ is then of the order $\sqrt{\frac{g_{\rm YM}^2 N}{J^2}}$
\footnote{
This fact signifies the strong non-locality of
the interaction. For local interaction, we would have
Neumann coefficients (for zero modes) of order $1$.
} \cite{RBAsymptoticNeumann},
we have $\Delta N^{\IIsII}_{00}\sim \sqrt{g_{\rm YM}^2 N} >\!>1$.
Now, 
the leading contribution 
will come from terms 
for which the factors $(\Delta N^{\IIsII}_{00})^{\frac{n}{2}}$
and $\frac{n}{2}!$ are comparable,
that is, terms with $n \sim \sqrt{g_{\rm YM}^2 N}$.
Although $\sqrt{g_{\rm YM}^2 N}$ is  large,
it is much smaller than $J$ in the regime we are working. 
Thus the use of pp-wave approximation is validated.

Performing the summation presents some interesting features.
Firstly, $n=0$ in the first term seems, at first sight, to make 
the only leading contribution because of the small denominator
$\Delta_\I+\Delta_{\II}-\Delta_{\III} \sim O(\frac{g_{\rm YM}^2 N}{J^2})$.
However, the numerator $E$ is of the same order
by non-trivial cancellation, hence
the contribution from the second term should also be evaluated.
Also other contributions from the first term become subleading.
Since the summation of the second term starts with $\frac{n}{2}=1$, 
the denominator can be replaced with $n$ to the leading order.
Then, the infinite series sums up essentially to the exponential function
except for the missing $n=0$ term,
$F e^{\Delta N^{\IIsII}_{00}} - F$.
Now, the exponent is large quantity  ($\sim \sqrt{g_{\rm YM}^2 N}$)
with negative sign, since $N^{\IIsII}_{00}<0$.
\footnote{
The sign differs from the literature since
our convention (\ref{RFladderAdSapprox}) of the $a^\dagger$ includes
a factor of $i$.}
Therefore the first term in this expression 
is extremely small and should be neglected in our approximation. 

Thus finally we have found
\beq
C\approxx \frac{E}{\Delta_\I+\Delta_{\II}-\Delta_{\III}} -F. \label{RFCisEF}
\eeq
For the three string vertex given in \cite{RBSpVol},
we get
using the asymptotic form of Neumann coefficients \cite{RBAsymptoticNeumann}
\beqa
E&=&
\frac{R^4}{J_{\III}^2 {\alpha^\prime}^2}
\frac{(\sin{m_\III \pi \rat})^2}{\pi^2 \rat}
\L(
-2 \delta^{(\alpha_\I \alpha_\III}
\delta^{\beta_\I) \beta_\III}
+
\frac{1}{2}
\delta^{\alpha_\I \beta_\I}
\delta^{\alpha_\III \beta_\III}
\R)
\non\\
F&=&
\frac{(\sin{m_\III \pi  \rat})^2}{\pi^2 \rat}
\Bigg(
\frac{-4 m_\III (m_\I/\rat)}{\L(m_\III^2-(m_\I/\rat)^2\R)^2}
\delta^{[\alpha_\I \alpha_\III}
\delta^{\beta_\I] \beta_\III}
-2
\frac{m_\III^2+ (m_\I/\rat)^2}{\L(m_\III^2-(m_\I/\rat)^2\R)^2}
\delta^{(\alpha_\I \alpha_\III}
\delta^{\beta_\I) \beta_\III}
\non\\
&&\qquad\qquad
-\frac{1}{2}
\frac{m_\III^2+ (m_\I/\rat)^2}{\L(m_\III^2-(m_\I/\rat)^2\R)^2}
\delta^{\alpha_\I \beta_\I}
\delta^{\alpha_\III \beta_\III}
\Bigg), \label{RFEandF}
\eeqa
where $\rat=J_\I/J_\III$.
Anti-symmetric and traceless symmetric pieces are
denoted by
\beqan
\delta^{[\alpha_\I \alpha_\III}
\delta^{\beta_\I] \beta_\III}
&=&
\frac{1}{2}(
\delta^{\alpha_\I \alpha_\III}
\delta^{\beta_\I \beta_\III}
-
\delta^{\beta_\I \alpha_\III}
\delta^{\alpha_\I \beta_\III}
), 
\\
\delta^{(\alpha_\I \alpha_\III}
\delta^{\beta_\I) \beta_\III}
&=&
\frac{1}{2}(
\delta^{\alpha_\I \alpha_\III}
\delta^{\beta_\I \beta_\III}
+
\delta^{\beta_\I \alpha_\III}
\delta^{\alpha_\I \beta_\III}
)
-\frac{1}{4}\delta^{\alpha_\I \beta_\I}\delta^{\alpha_\III \beta_\III}
,
\eeqan
respectively.
Substituting (\ref{RFEandF}) into (\ref{RFCisEF}), we see that
the the gauge theory results \cite{RBGauge1}\cite{RBGauge2} are reproduced. 

At this point, let us  clarify the issue of
ambiguities of the three string vertex.
We consider unitary transformations,
$
H_{free}+H^\prime_{int} = (1+D+\ldots)^{-1}(H_{free} + H_{int})(1+D+\ldots),
$
or,
\beq
H_{int}^\prime=H_{int}+[H_{free}, D], \label{RFUnitarytrans}
\eeq
where $H_{int}$ and $D$ have  the order of the string coupling constant.
$H_{int}^\prime$ has the same
symmetry as the original $H_{int}$ by construction. 
In particular, it satisfies the constraints of supersymmetry.
These transformations
can be considered as field redefinitions, of which $D$'s are the generators. 
In usual field theory, it is guaranteed that the physical
observables do not change under these redefinitions, provided
the locality of the transformations.
We can actually show that 
also in our case the observable $C$ does not depend
on the transformation (\ref{RFUnitarytrans}) for a broad
class of $D$.
Indeed, for any $D$ which is given by 
the overlap part with a polynomial prefactor,
$|D\rangle=P(a^\dagger) e^{\frac{1}{2} a^\dagger N a^\dagger}|0\rangle$,
we get from (\ref{RFTheRulepp})
\begin{equation}
C^\prime=C+
\langle 0| e^{-\sqrt{2 \Delta} \frac{x_2^\mu}{|x_2|} a^\mu } |D\rangle=C+
P\L(-\sqrt{2 \Delta}\frac{x_2^\mu}{|x_2|}\R) e^{ \Delta N^{\IIsII}_{00}}.
\end{equation}
Thus, $C$ does not change up to exponentially negligible term,
which should be neglected in our approximation.
Hence, our method does not suffer from these ambiguities.
\footnote{
This property makes it natural
to call these observables as on-shell,
although $\Delta_1+\Delta_2\neq\Delta_3$.
This is also natural from the viewpoint of the holographic ansatz
since they are given by the path-integral with a fixed 
asymptotic behaviour of the fields.
We note that for ordinary transition amplitudes
in time-independent systems, it is the condition
$\Delta_1+\Delta_2=\Delta_3$
which guarantees that
they do not change upon field redefinitions.
}
The form of $D$ given above closely resembles that of the 
generator of a local field redefinition 
in ordinary field theory, the overlap part and the polynomial of $a^\dagger$'s 
respectively corresponding to the $\delta$-function part and the polynomial of derivatives.

A special case of (\ref{RFUnitarytrans}) is the
three point vertex given in 
\cite{RBRedefVertex}, which can be written as
$[H_{free}, D]$ with $D$ the overlap part itself.
For the vertex, $E$-term and $F$-term in (\ref{RFCisEF}) 
cancel so that it gives a null contribution to $C$, as expected from the above argument. 
Previously, the following relation has been proposed\cite{RBppPrevRule}
\beq
C=\frac{E}{\Delta_\I+\Delta_{\II}-\Delta_{\III}}, \label{RFprevrule} 
\eeq
which lacks the $F$-term, so that the results of which are affected by the ambiguities.
Recently, Dobashi and Yoneya
\cite{RBDY}
\cite{RBDY2}
have succeeded in 
reproducing the gauge theory results
using (\ref{RFprevrule})
(with some refinements) by choosing a 
vertex which is a special linear combination of the 
vertices of \cite{RBSpVol} and \cite{RBRedefVertex}.
\footnote{
Bosonic part of this vertex is first discussed in \cite{RBChuKhoze}.
Very recently, the work \cite{RBLeeRusso} appeared along this line.}
In our perspective, their success follows from the fact that
$F=0$ holds for their choice. 

We should be able to perform many tests of the AdS/CFT correspondence 
by further applying the framework proposed in this letter.
It will be interesting to
study (a)
the subleading order of
our approximation which has rather
intricate structures, (b)
general BMN operators such as
those which have vector impurities,
for which one should use different propagators and 
$x_r$-dependences (\ref{RF3pf}), (c)
the four (or more) point functions.

The author would like to thank 
T. Yoneya, S. Dobashi for many discussions and encouragements.
The author also would like to thank
T. Sato, Y. Aisaka, K. Hashimoto,
T. Muramatsu,
T. Shimada for discussions and encouragements.
The author would like to thank 
particle theory group at YITP and Kyoto university
for their hospitality during the author's stay at YITP.
The author would like to thank S. Iso, S. Sekino, N. Sasakura,
and H. Kawai for useful comments.
This work is supported in part by JSPS Research Fellowships
for Young Scientists (\#16-10606).

\end{document}